\input{aipcheck}

\documentclass[
    ,final            
  ]
  {aipproc}

\layoutstyle{6x9}

\begin{document}

\title{Collisional Grooming of Debris Disks}


\author{Marc J. Kuchner}{
  address={NASA Goddard Space Flight Center, Exoplanets and Stellar Astrophysics Labratory, Greenbelt, MD 20771}
}

\author{Christopher C. Stark}{
  address={University of Maryland, Departent of Physics, College Park, MD 20742}
}

\begin{abstract}

Debris disk images show clumps, rings, warps, and other structures, many of which have been interpreted as perturbations from hidden planets. But so far, no models of these structures have properly accounted for collisions between dust grains. We have developed new steady-state 3D models of debris disks that self-consistently incorporate grain-grain collisions. We summarize our algorithm and use it to illustrate how collisions interact with resonant trapping in the presence of a planet.

\end{abstract}

\maketitle


\section{Introduction}

Debris disk images have emerged as important tools for finding extrasolar planets. A 2005 image of Fomalhaut's debris disk \citep{kala05} showed a ring offset from the central star pointing to a hidden planet; two years later, new \textit{HST} Advanced Camera for Surveys (ACS) images revealed one of the first-ever directly imaged exoplanets orbiting inside the disk \citep{kala08}. The new generation of high contrast coronagraphs (NICI, HiCIAO, SPHERE, etc.) and mid-infrared/submillimeter imaging tools (\textit{JWST}, SOFIA, ALMA) promises to return even more spectacular images of disks and planetary systems. 
But we cannot yet reliably decode the spectacular images of debris disks made by \textit{HST}, JCMT, and other observatories, partly because no current models of structures in debris disks properly account for dust grain collisions. Many of the structures we see probably arise from the interaction between planetary perturbations, grain-grain, and grain-rock collisions. So even for signatures like the offset in the Fomalhaut ring that seem to be clear planet signposts, we have not yet been able to reliably derive planet orbital parameters.

The key parameter that shows the relative importance of collisions for a population of grains in a debris disk is the ratio of the collision time, $t_{coll}$, to the Poynting-Robertson (PR) time,  $t_{PR}$, the time it takes the grains to spiral into the star under PR drag \citep{burn79}. This ratio, $\eta = t_{PR}/t_{coll}$, represents roughly the number of collisions a particle undergoes during its lifetime in the absence of resonant trapping (which can increase the particle's lifetime by an order of magnitude.)  The collision time, in turn, is approximately set by the cloud's face-on optical depth,  $\tau_\perp$;  $t_{coll}\approx t_{Orbit}/12 \tau_{\perp}$ where $t_{Orbit}$  is the orbital time \citep[e.g.][]{wyat05}. For typical imaged debrs disks (e.g. Vega, $\beta$ Pictoris, Fomalhaut, etc.) $\eta$ is roughly 0.1--10 for a 10 $\mu$m grain. 

We describe here a new kind of debris disk model that self-consistently incorporates grain-grain collisions. Our algorithm yields a dust distribution that simultaneously solves the governing dynamical equations and the mass flux equation in 3D. It uses the 
results of a collisionless disk simulation as input and iteratively "grooms"
the disk until it converges on the correct density distribution for a collisional disk, allowing us to model disks with $\eta$ up to at least a few hundred. 

\begin{figure}

 \includegraphics[height=10 cm, angle=90]{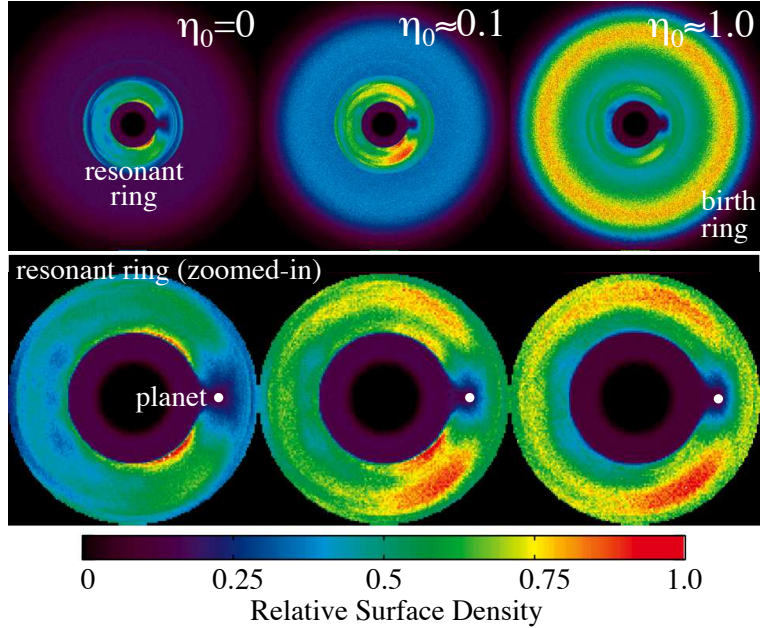}

  \caption{Surface density as a function of $\eta_0$ for a disk with an Earth-mass planet on a circular orbit at 1 AU, no fragmentation. The top row shows the entire disk, which extends to 4.25 AU. The bottom row shows zoomed-in views of the resonant ring. Collisions emphasize the birth ring and change the shape of the resonant ring. }
\end{figure}
  
\section{Numerical Techniques}
  

To model the structures of dust clouds containing planets, we begin by constructing a "seed model". We numerically integrate the orbits of dust particles, a planet, and a star, as many authors have done 
\citep{derm94, wiln02, moro02, mora04, dell05, wyat06, star08}. To the gravitational forces on the particle we add PR drag, radiation pressure, and solar wind drag. Radiation pressure effectively reduces the mass of the star as seen by the particle by the factor $(1-\beta)$. In the solar system, $\beta$ is roughly $(0.285 \mu \rm{m})/s$ for a spherical particle with radius $s$ and density 2 g cm${}^{-3}$. To model solar wind drag, we modify the PR drag coefficient, as many of the above authors have done.
We release the particle in an orbit representing that of an asteroid or other parent body. The particle jumps to a new 2-body orbit appropriate for the radiation pressure it feels, conserving its velocity at release. We follow the particle until it moves outside our region of interest or collides with the Sun or a planet. At regular intervals, we record the location of the particle. We use the recorded locations to produce a 3D histogram modeling the steady-state distribution of particles released at regular intervals from a belt of small bodies.  


Our simulations use many more particles than the previous generation of models. This capability is important because the density distributions produced by the standard techniques suffer from three key sources of uncertainty that can be alleviated using larger numbers of particles: Poisson noise in each cell of the histogram, Poisson noise in the population of the mean motion resonances (MMRs) and the domination of the histograms by a few long-lived particles \citep[e.g.][]{moro02}. 
Our models are the first to properly manage all three sources of noise. We handle Poisson noise by using at least 5,000 particles per size bin, choosing an appropriate interval for recording particle positions and by devoting equal amounts of computer time to particles with large and small $\beta$ values. Our collision algorithms (discussed below) manage the third source of noise by destroying long-lived particles. 



To make the most of the available computers, we developed our own hybrid symplectic integrator adapted for the dust problem. This integrator combines a Burlisch-Stoer integrator and a symplectic integrator \citep{wisd91} in the manner of \citep{cham99}; the Bulirsch-Stoer integrator turns on during close encounters. We have been running our codes on NASA's Discover cluster (2,560 processors: 512 3.2 GHz plus 2,048 2.6 GHz). We maintain a free online database of our collisionless debris disk simulations at

\url{http://asd.gsfc.nasa.gov/Christopher.Stark/catalog.php}.

Once the seed model is finished, we apply our new algorithm for incorporating grain-grain collisions. We use the following iterative approach:
\begin{enumerate}
\item Integrate the orbits of the particles as described above (the seed model).
 
\item Record both the positions and the velocities in a 6-dimensional histogram: a distribution function. 

\item Allow the particles to interact with the distribution function assembled in steps 1 and 2; update the distribution function. 

\item  Repeat step 3 until the process converges.
\end{enumerate}
Using the assumption that all collisions lead to complete, instantaneous grain removal \citep{back93}, we have produced the first 3D models of exozodiacal clouds that incorporate collisions in steady-state cloud self-consistently.

In this picture, we view each of the "particles" in the simulation as particle streams representing billions of particles each. As they progress through the cloud, the streams become attenuated by collisions with other streams as $e^{-\tau}$, where $\tau$ is the optical depth of the cloud along the stream; $\tau = \int <n\sigma> ds$. Between any two records in the histogram, we approximate the optical depth for collisions along the stream $i$ as  $\Delta \tau \approx \sum_j n_j \sigma_j|\vec v_i - \vec v_j| \Delta t$, where $n_j$, $\sigma_j$ and $\vec v_j$ are the number density, particle cross section, and vector velocities of all the streams that flow through a given histogram cell, as measured in that cell, and $\Delta t$ is the time interval between the two records. This approximation works as long $\Delta t << t_{coll}$, which is always true for our simulations.
We use the iteration scheme described above to make sure that all the streams are attenuated consistently with one another. We examined five possible iterative schemes that achieve self-consistency. Two that met our requirements for accuracy; we chose the faster one. This scheme allows us to process a simulation containing 5,000 particle streams in 15 minutes on a single processor. The algorithm typically converges in $\sim 5$ iterations to a point where the solution found in each subsequent iteration differs by no more than 1\% from the previous iteration's solution in any cell. The time it takes to run a 5,000-particle simulation with our algorithm is dominated by the time it takes to integrate the orbits of the particles, not the collision algorithm. 

To test our simulations, we compared them to analytic solutions to the azimuthally-symmetric steady-state mass flux equation for dust grains under PR drag assuming that collisions serve only to completely remove particles \citep{wyat05}. We found that our simulations disagree substantially with the analytic solutions only for  $\eta_0 > 235$; for such short collision times, the histogram bins become too small to resolve the radial structure. To prevent this problem, we must simply be careful to choose the bin size,  $d$, so that $d<r_0/(2\eta_0)$. 


\section{Results and Discussion}

{\bf Figure 1} shows some examples of our first generation of collisional disk models with no daughter particles. It shows the same disk/planet combination with three different values of $\eta$. An Earth-mass planet in this simulation (semi-major axis 1 AU) traps some dust in MMRs, creating a resonant ring.
An emerging common description of debris disks is that they have three radial regions \citep{stru06}
  \newcounter{Lcount}
  \begin{list}{\Roman{Lcount}.}
    {\usecounter{Lcount}
    \setlength{\rightmargin}{\leftmargin}}
\item A ``birth ring'' of large bodies, perhaps analogous to the asteroid belt or Kuiper belt.
\item A region interior to the birth ring dominated by large grains spiraling in under PR drag and
\item A skirt of small unbound or marginally-bound grains ($\beta$-meteoroids) exterior to the birth ring. 
 
\end{list}
 
The simulation shown in {\bf Figure 1} models regions I and II. As $\eta$ increases, the birth ring becomes emphasized with respect to the resonant ring, which is in region II. The resonant ring also changes shape as $\eta$ increases; sharp structures in the inner part of the ring vanish, and the azimuthal structure shifts. Clearly, understanding these effects will be important for understanding images of debris disks and probably our own zodiacal cloud. 
{\bf Figure 2} shows the collision rate as a function of semi-major axis in one of our collisional models (green). The collision rate is enhanced within the MMRs. The effect is even more pronounced in the disk images than this plot might suggest, because the particles spend most of their time in the MMRs. The first-order MMRs show double-peaked profiles, probably associated with their large libration amplitudes. An orange dashed curve shows the approximation, $t_{coll} \approx t_{Orbit}/12 \tau_{\perp}$  used in previous work.

So far, these results mostly illustrate the potential of this algorithm. We are working on new collisional models that incorporate daughter particles, a range of particle sizes, and a range of parent body dynamics. We expect that these kinds of models will soon help answer some important questions in debris disk modeling: How robust are the structures predicted by 3-body simulations?  How does the distribution of parent bodies impact the appearance of the dust cloud?  They should be useful for recognizing planet migration in debris disks and providing better constraints on the orbital parameters of planets in eccentric rings like Fomalhaut.

\begin{figure}
\includegraphics[height=9.2 cm, angle=90]{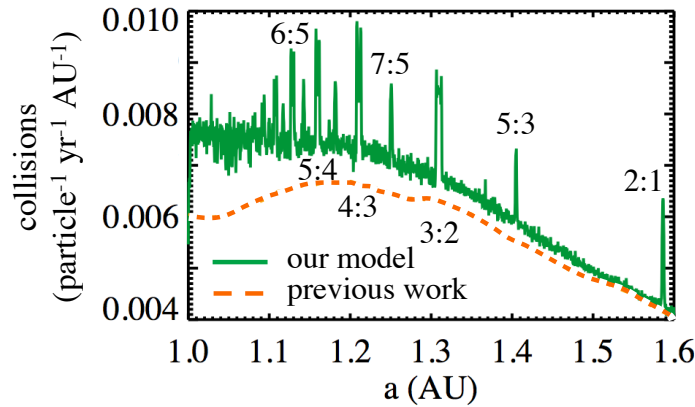}
  \caption{Collision rate in one of our simulations vs. semimajor axis (green) compared to the naive estimate sometimes used in previous work (orange dashed).}
\end{figure}


\begin{theacknowledgments}

We thank NASA's Astrobiology program, NASA's Graduate Student Researchers Program and the NASA High-End Computing Program. 

\end{theacknowledgments}



\bibliographystyle{aipproc}   

\end{document}